\documentclass[]{spie}
\usepackage[dvips]{graphicx}

\title{High Contrast Imaging with Gaussian Aperture Pupil Masks}
\author{John H. Debes, Jian Ge, Caylin Mendelowitz, and Anne Watson
\skiplinehalf
Pennsylvania State University, 525 Davey Lab, University Park, USA
}

\begin{document}
\maketitle
\begin{abstract}
Gaussian aperture pupil masks (GAPMs) can in theory achieve the contrast
requisite for directly imaging an extrasolar planet.  We outline the process
 of fabricating and testing a GAPM
for use on the Penn State near-IR Imager and Spectrograph (PIRIS) at the Mt. 
Wilson 100$^{\prime\prime}$ telescope.  We find that the initial prototype
observations are quite successful, achieving a contrast similar to a traditional Lyot coronagraph without blocking any light from a central object
and useful for finding faint companions to nearby young solar analogues.  In 
the lab we can reproduce the expected PSF to within an order of magnitude
and with new designs achieve $\sim5 \times\ 10^{-5}$ contrast at 10$\lambda/D$.
  We find that small inaccuracies in the mask fabrication process and 
insufficient correction of the atmosphere contribute the most degradation to 
contrast.  Finally we compare the performance of GAPMs and Lyot coronagraphs
of similar throughput.
\end{abstract}

\keywords{high contrast imaging, apodization, extrasolar planets, coronagraphy}

\section{Introduction}
\label{intro}
The search to directly image an extrasolar planet requires contrast
levels of $\sim$10$^{-9}$ a few $\lambda / D$ from the central star.
Scattered light in a telescope
and the diffraction pattern of the telescope's aperture limit the contrast
possible for direct detection of faint companions.  The circular aperture of
telescopes creates a
sub-optimal diffraction pattern, the so-called Airy Pattern which is
azimuthally symmetric.  In addition, the intensity in the diffraction pattern
of the circular
aperture declines as $\theta^{-3}$.  Currently the best way to diminish
the Airy pattern is to use a coronagraph by using the
combination of a stop in the focal plane that rejects a majority of the
central bright object's light and a Lyot stop in the pupil plane to reject high
frequency light \cite{lyot,malbet96,sivar01}.  Several recent ideas explore the use of alternative ``apodized'' apertures for high contrast
imaging in the optical or near-infrared \cite{nisenson01,spergel01,ge02a}.
These designs revisit concepts first experimented with in the field of optics
\cite{jacquinot64}.  Other designs, such as the band
limited mask, seek to null the light from a central star in much the same way 
that a nulling interfermoeter performs \cite{traub}.  

All of these designs in theory can
reach the contrast ratio necessary for imaging a planetary companion, however
most of these designs have yet to be tested in the lab or on a real telescope
where other concerns arise.  The specific advantages of each idea cannot be 
determined until they are actually built or modeled in such a way as to 
simulate real engineering problems. 

A promising design for a shaped aperture was recently suggested
by Ref.~\citenum{spergel01}.  In this case the top and bottom edges of the aperture
are described by gaussian functions:
\begin{eqnarray}
y_t & = & a R\left\{ \exp \left[-\left(\alpha x/R\right)^{2}\right] -\exp \left(-\alpha^{2}\right) \right\} \\
y_b & = & -b R\left\{\exp\left[-\left(\alpha x/R\right)^{2}\right] -\exp\left(-\alpha^{2}\right) \right\},
\end{eqnarray}

where $x$ goes from $-R < x < R$.  The Fourier transform of the aperture
shape function gives the resulting
diffraction pattern in the imaging plane.  Since the Fourier transform of a
gaussian is another gaussian, the intensity of the diffraction pattern 
$I(\epsilon, \eta)$ along
one image plane axis decreases exponentially, which we denote the high contrast
 axis.  The ratio $z=I(\epsilon,\eta)/I(0)$ gives the contrast relative to the
peak intensity of the diffraction pattern.  The variables $a, b,$ and $\alpha$
 are all free parameters that can be
used to optimize the aperture for depth of contrast, the angle from the central
object at which high contrast starts, and the azimuthal area of high contrast.

\begin{figure}
   \begin{center}
   \begin{tabular}{c c}
   \includegraphics[height=6cm]{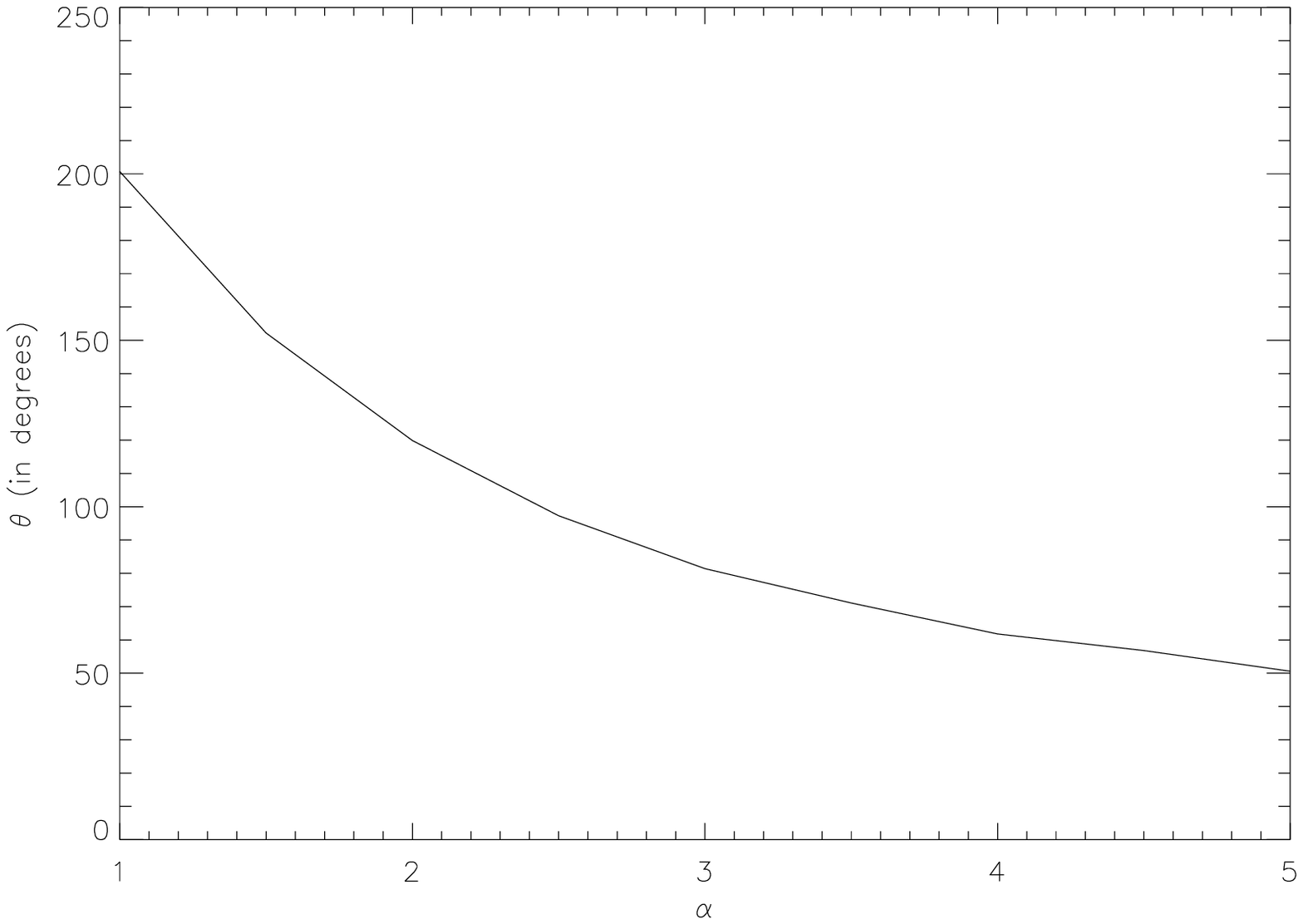} & 
   \includegraphics[height=6cm]{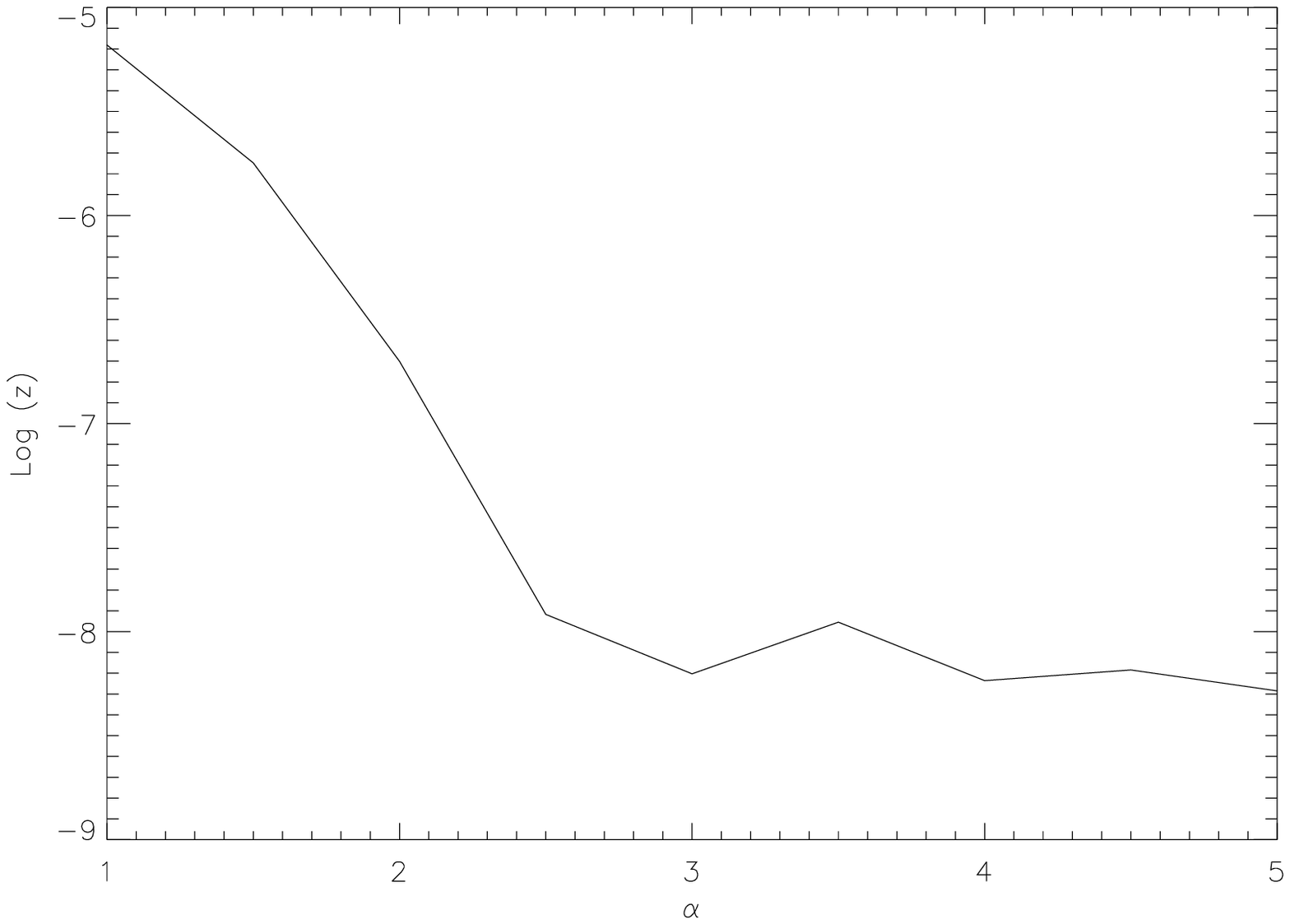}
   \end{tabular}
   \end{center}
   \caption[example]
{ \label{fig1} (left) The log of normalized contrast vs. $\alpha$.  As $\alpha$
increases the contrast ratio decreases sharply, but at the cost of where the
contrast begins.  (right) The dependence of search angle $\theta$ with 
increasing $\alpha$.  A larger $\alpha$ restricts the total region of high 
contrast.
}
   \end{figure}

By placing a mask into the pupil plane with a gaussian aperture, one can
transform a traditional circular aperture telescope into one with a diffraction
pattern better suited for high contrast imaging.  Using a mask represents a
quick, efficient, and cheap way to test this emerging imaging method to 
determine its advantages and tradeoffs and compare them to the performance of
other existing techniques.

We have endeavored to begin anwering the question of which design ultimately 
will be useful in the search for extrasolar planets, or which will be most
useful for other areas of astronomy where less stringent tolerances are present.
To that end we have designed, fabricated, and tested several GAPM designs for
use with the Penn State near-IR Imager and Spectrograph (PIRIS)\cite{ge02a}.  In Section 
\ref{design} we explore what the best design for a telescope would be.  In 
Section \ref{fab} we briefly discuss the process of fabrication of the GAPMs,
while in \ref{test} we discuss the various tests we performed in the lab and
on the ground at the Mt. Wilson 100$^{\prime\prime}$ telescope.  In Section
\ref{equal} We compare the contrast of GAPMs with comparable Lyot coronagraphs
Finally in section \ref{concl} we discuss
what role GAPMs have for future high contrast imaging.
  
\section{Designing a GAPM for current telescopes}
\label{design}
The idealized design of a single gaussian aperture in practice cannot be used
on current telescopes due to their circular secondary obstructions and the 
presence of the support structure.  These two additions serve to modify the 
resulting diffraction pattern and destroy the advantages of the single 
aperture.  Therefore, a new design that avoids or minimizes their effect is 
necessary to retain high contrast.  In order to design a GAPM for use on the
Mt. Wilson 100'' telescope, we performed several numerical simulations to 
determine the design that had the best theoretical performance.  In addition we
quantify how the presence of a secondary mirror and support structure degrade
contrast.

 \begin{figure}
   \begin{center}
   \begin{tabular}{c c}
   \includegraphics[height=6cm]{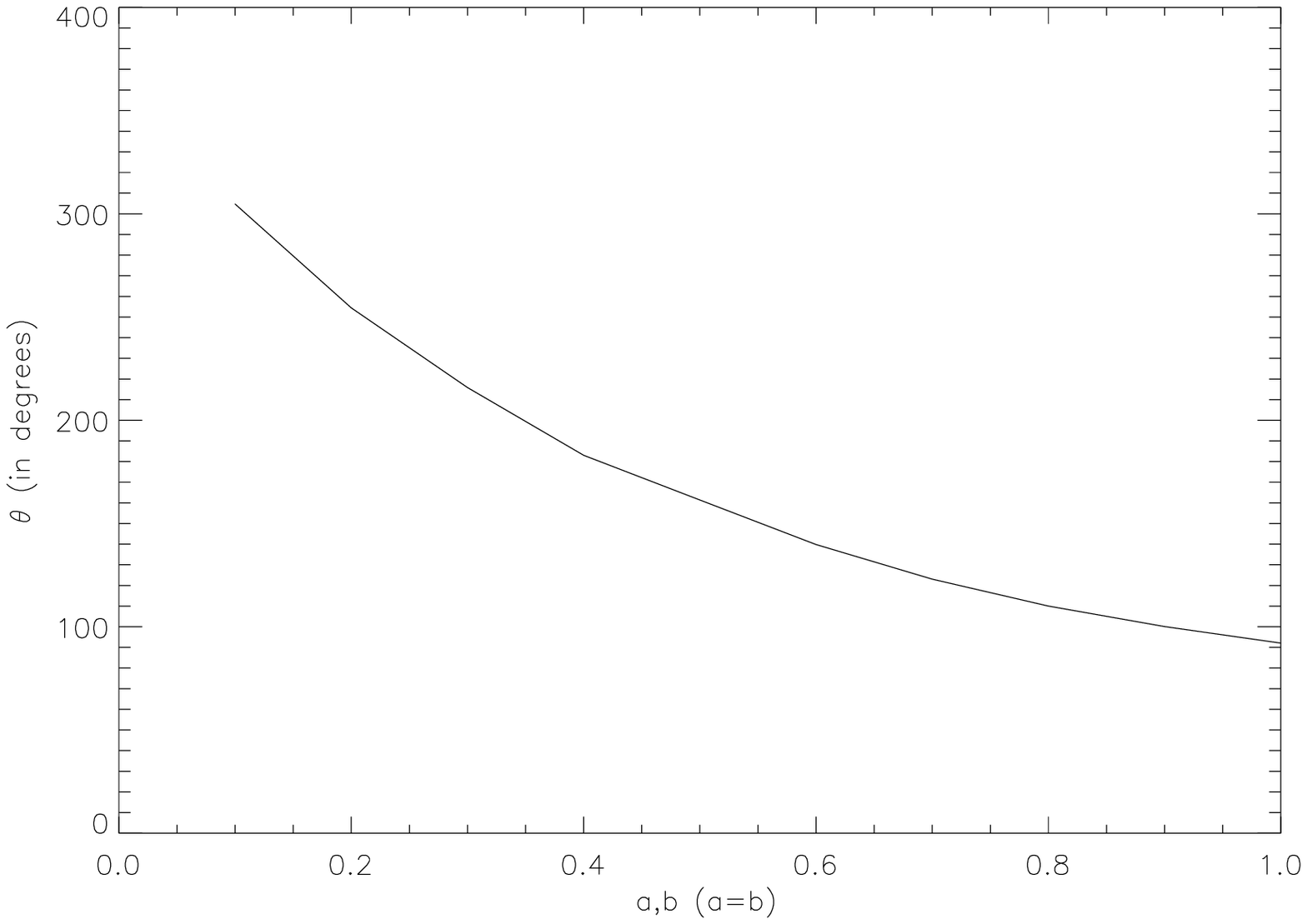} & 
   \includegraphics[height=6cm]{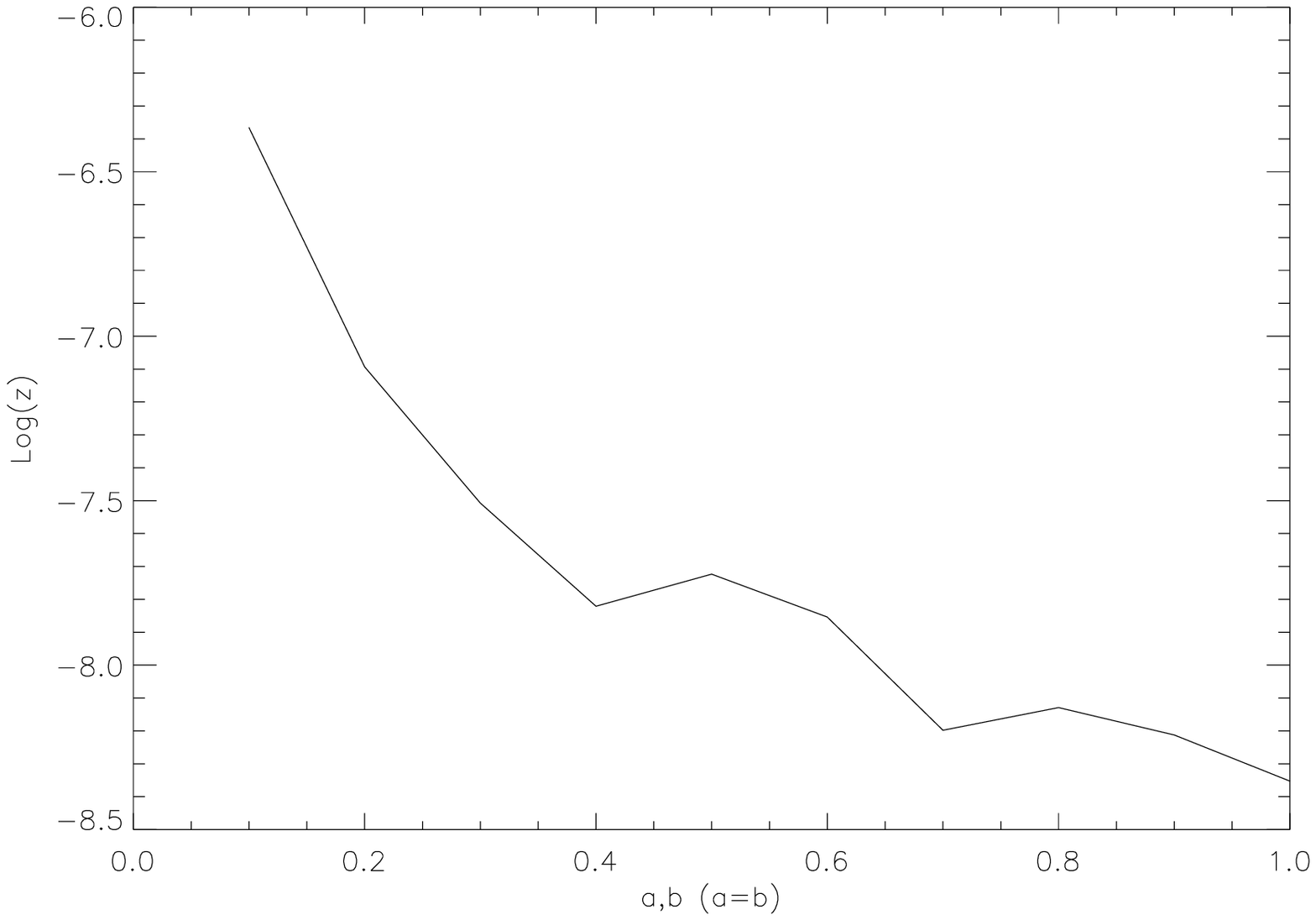}
   \end{tabular}
   \end{center}
   \caption[example]
{ \label{fig2} Same as for Figure \ref{fig1} but for varying $a,b$ where $a=b$.  
Decreasing $a$ or $b$ can increase search angle but decrease throughput and
degrade contrast.
}
   \end{figure}

\subsection{Modeling Different Apertures}

In order to test the performance of different designs, we have created several
IDL programs to model the resulting diffraction patterns or point spread
functions (PSFs) from different 
apertures.  In its simplest form, the models determine the Fraunhofer 
diffraction of light through an aperture.  The diffraction is calculated by
performing a Fast Fourier Transform (FFT) algorithm on an incoming wavefront, 
represented by a complex 2D array.  Taking the real part of the resultant 
amplitude image
and squaring it gives the intensity.

The complete model calculates the diffraction of a light wave's amplitude
first passing through the 
circular aperture of a telescope with a secondary mirror and support structure.
An input wavefront array with the shape of the pupil undergoes an FFT to give
a PSF of the telescope.  After calculating the PSF of the 
aperture, an ideal test image can be convolved with the PSF.  This allows an
independent calculation of the contrast possible for a given design by 
creating an image of a central object and a companion of a given flux ratio.  A crucial
issue for any design is its ability to not only achieve a high contrast, but
do so in a way that does not seriously degrade the intrinsic resolution of the
companion PSF.  An example would be an overly ambitious undersizing of a Lyot
 stop.  
While in theory this can provide very high contrast, the smaller aperture
size essentially spreads out the light of any faint companion and drastically 
increases the integration time, making it harder to detect.

After convolution with the image, the array is then multiplied with the 
transmission of a focal plane mask, which can either be empty or have some 
coronagraphic spot present.  Next the array undergoes an inverse FFT which results
in the intensity pattern at the pupil plane.  A pupil mask transmission 
function is then multiplied with the array and udergoes one final FFT to create
the image on the detector.  In order to have sufficient resolution and 
keep aliasing affects to a level of $<$10$^{-8}$, most simulations were 
performed with 2048 $\times$ 2048 element arrays where the pupil was undersized
by a factor of two.  The model does not include wavefront errors, residual 
scattered light, or atmospheric turbulence.  A single wavelength PSF is 
generated.  

\subsection{Testing Initial Designs}

Studying the effect of varying $a$,$b$, and $\alpha$ is instructive in 
understanding the best values to use for a given situation.   For the initial
test of a prototype design, a balance of search area and depth of contrast was
most useful.  Figure \ref{fig1} shows the effect of varying $\alpha$ on 
$z$ along the axis of highest
contrast.  Around $\alpha$=2.7 one can see that the contrast doesn't improve 
due to the resolution of the simulations.  To determine $z$ we took
ten points centered at $\sim 10-20 \lambda/D$ depending on where the region
of high contrast occurred and averaged their value of
 $z$.  In general,
increasing $\alpha$ will decrease $z$, but at the 
cost of the region of high contrast starting at a larger $\lambda/D$.

  Another important factor is 
determining the azimuthal coverage of the region of high contrast.  we measured
this quantity, $\theta$ by calculating the region that had  $z$ $<$ 10 $z_{hc}$
where $z_{hc}$ is the contrast ratio along the high contrast axis at the.
  Figure 
\ref{fig1} shows the variation of $\theta$ with $\alpha$.  As $\alpha$ 
increases, $\theta$ decreases.
\begin{figure}
   \begin{center}
   \begin{tabular}{c}
   \includegraphics[height=8cm]{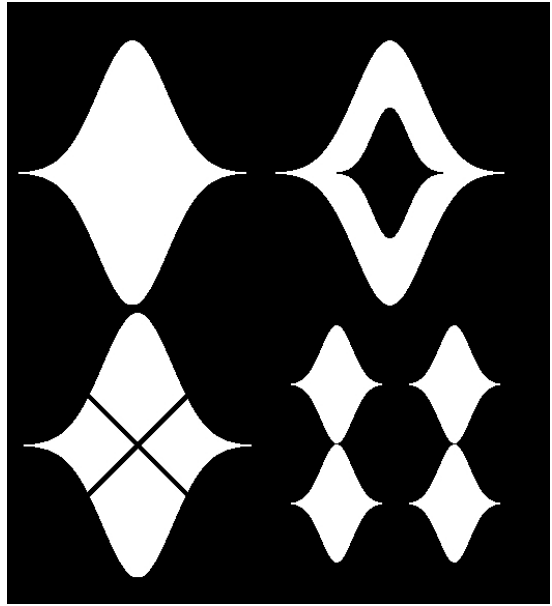} 
   
   \end{tabular}
   \end{center}
   \caption[example]
{ \label{fig3} The designs tested in our simulations.  The upper left is the
single aperture design, upper right single aperture with secondary, lower left
single aperture with support structure, and lower right multiple aperture
design.
}
   \end{figure}

Varying $a$ and $b$ can also affect azimuthal coverage and contrast.  In 
general decreasing $a$ and $b$ has the affect of increasing azimuthal coverage
with the tradeoff of a reduced contrast.  As an example we have plotted
both $\theta$ vs. $a, b$ and $\log z$ vs. $a, b$ (see Figure \ref{fig2}).  
Decreasing $a$ and $b$ also 
reduces throughput.

As mentioned before, a single open aperture is degraded severely by the 
addition of a secondary and support structure.  It is then 
necessary to determine a design that avoids the 
problems associated with these structures.  Two types of designs are 
immediately apparent, either mitigating the effect of the secondary mirror
or avoiding the secondary and support structures altogether.

We created four designs to test with our simulations in order to gauge the 
degradation in contrast from the ideal case and to determine what configuration
was the best.  The four designs are shown in figure \ref{fig3}.  The first is 
an ideal case, a single gaussian aperture with $\alpha$=2.7 and $a=b=1$.  The
second design is a single gaussian aperture with a smaller gaussian stop that
blocks the circular secondary completely, but with no support structure.  A
third design has only a support structure, in order to disentangle the 
level of degradation of the two designs.  The orientation of the support structure is crucial, since the spider arms can completely ruin the high contrast 
axis if positioned perpendicular to the mask's horizontal axis.  
The best design
requires that the support structure is rotated 45$^\circ$ with respect
to the horizontal axis.  When that is done the resulting diffraction spikes are
arrayed in the brighter regions of the pattern.  The final 
design has four gaussian
apertures that are equal to half the total width of the single aperture and 
avoid any other structure.

\begin{figure}
   \begin{center}
   \begin{tabular}{c c}
   \includegraphics[height=8cm]{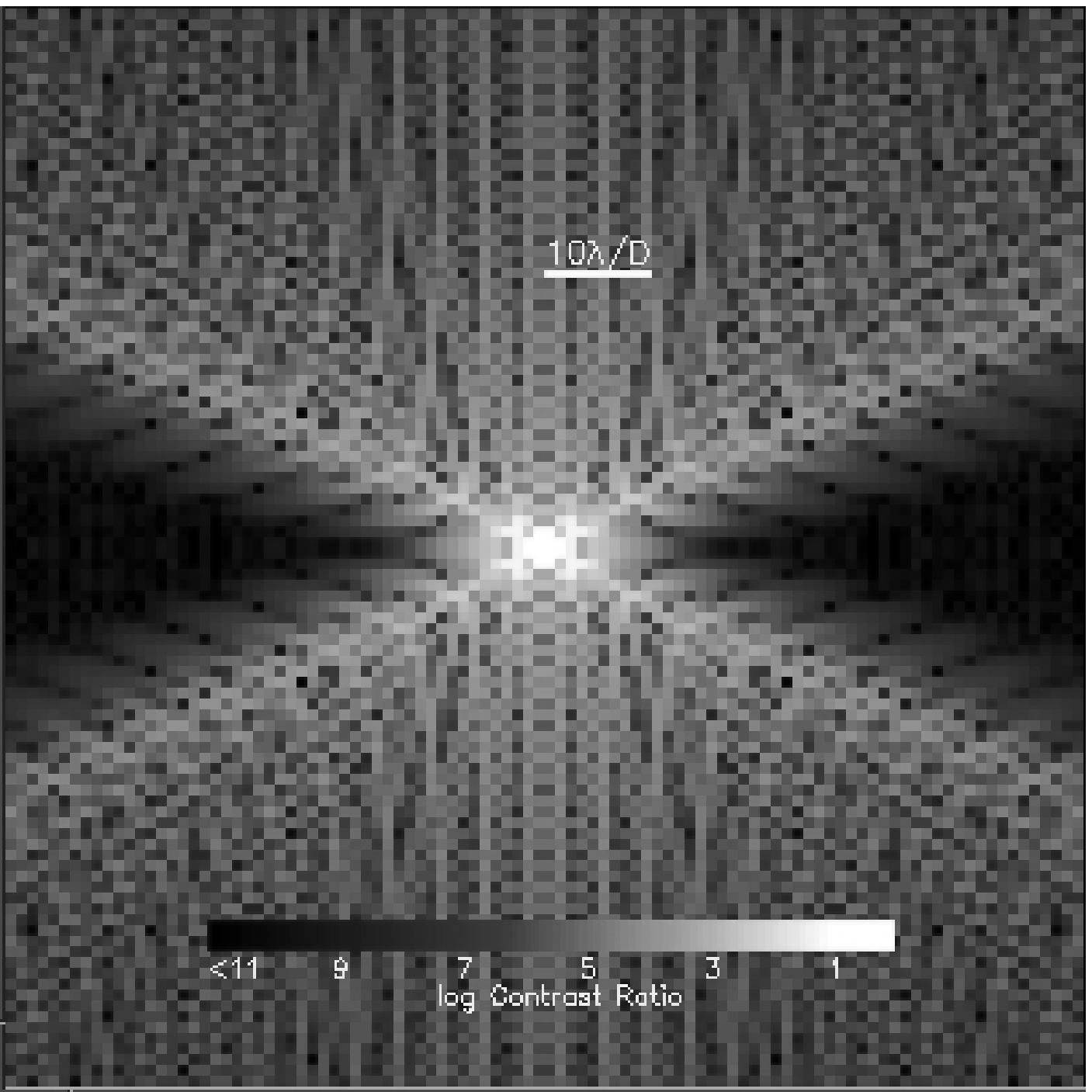} & 
   \includegraphics[height=8cm]{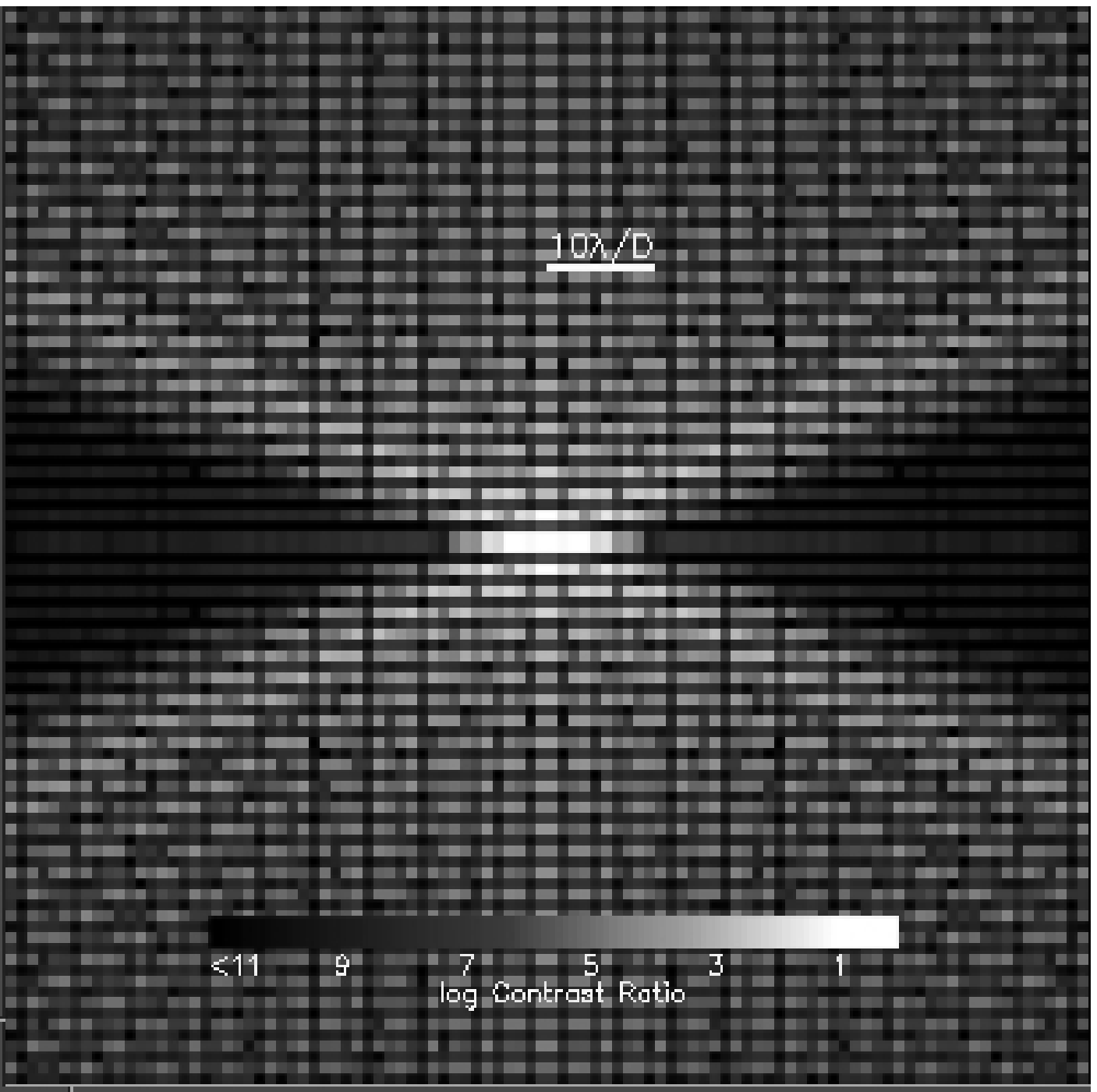}
   \end{tabular}
   \end{center}
   \caption[example]
{ \label{fig4} The resulting diffraction patterns for the single aperture with
secondary design (left) and the multiple aperture design (right).  The images
are logarithmically scaled.
}
   \end{figure}

Figure \ref{fig4} shows the results of two of our simulations, the multiple 
opening mask and the mask with a secondary and demonstrates the basic 
diffraction pattern of a gaussian aperture.  The peak is surrounded by two
different regions, a dark high contrast region around the horizontal axis
and a brighter low contrast region in an hourglass shape centered vertically 
from the peak.  The gaussian aperture effectively redistributes the light from
the central object, leaving a region with little light.

  Compared to the single opening
mask, the mask with a secondary has a much wider bright central region and a 
slightly smaller search area.  The four opening approach has an overall 
similarity to the single opening mask, but has an elevated background 
and less resolution due to the smaller opening.  In order to quantify the contrast, each design's ratio of intensity to the peak was averaged between 
10$\lambda/D$ and 20$\lambda/D$ along the high contrast axis.  This was 
performed over a large range of simulation array sizes to determine the
best array size.  

  Figure \ref{fig6} shows the full results and demonstrates the 
problems of adding a gaussian secondary and support structures.  The secondary,
while an improvement over a circular shape, still degrades contrast close to
the central object.  At larger distances the secondary has a contrast similar
to the single opening design.  The support structure for all the array sizes
was chosen to be .2\% of the primary size (the approximate ratio at the
Mt. Wilson telescope).  We find that this degrades contrast by
$\sim$2-3 orders of magnitude.  The multi-aperture 
design has slightly degraded contrast, and a major trade off is a twofold 
decrease in resolution due to the halving of the aperture width.  Even with
this degradation in resolution, it performs better than the gaussian secondary
and support structure combined.

\begin{figure}
   \begin{center}
   \begin{tabular}{c}
   \includegraphics[height=6cm]{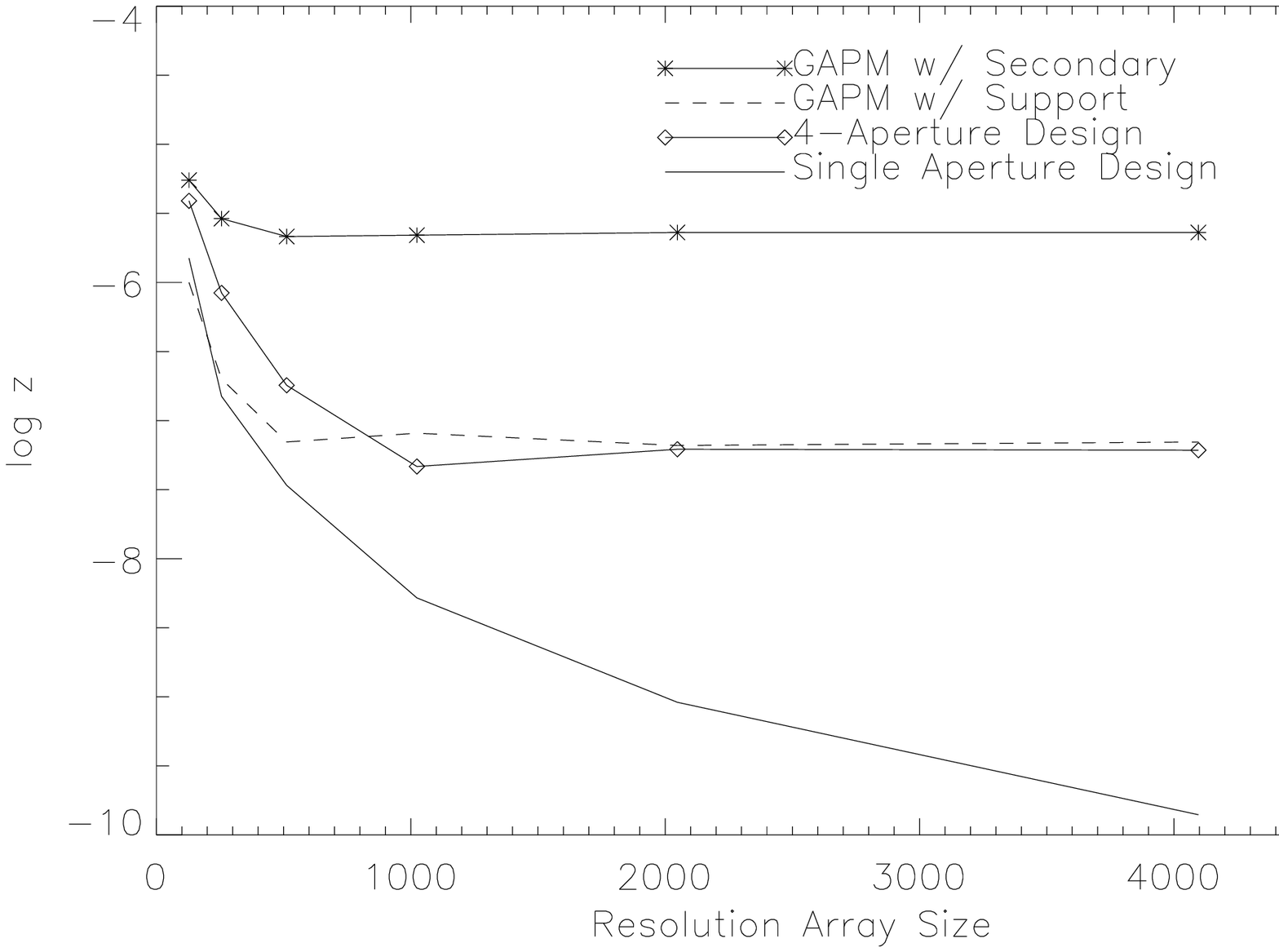} 
   \end{tabular}
   \end{center}
   \caption[example]
{ \label{fig5} Contrast as a function of simulation array size for different
types of designs.
}
   \end{figure}

\begin{figure}
   \begin{center}
   \begin{tabular}{c c}
   \includegraphics[height=7cm]{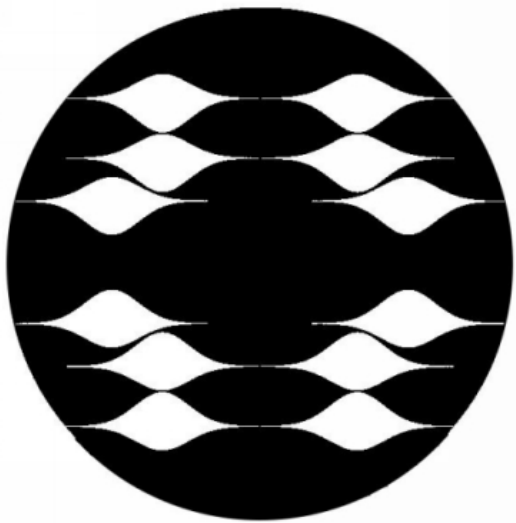} &
   \includegraphics[height=7cm]{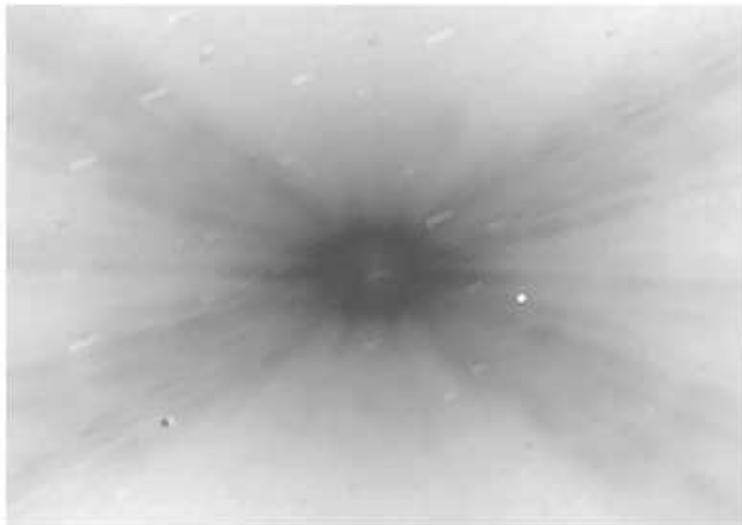}
   \end{tabular}
   \end{center}
   \caption[example]
{ \label{fig6} Picture of the final design and of the resulting diffraction pattern at Mt. Wilson telescope.
}
   \end{figure}

Our final design used the multi-aperture approach.  We chose $\alpha$=2.7,
$a=.23$ and $b=.33$ to give a $z$=10$^{-76}$ contrast and 180$^\circ$ 
azimuthal coverage.  Figure \ref{fig5} shows the design and the 
resulting diffraction pattern.  We arrayed 12 apertures in a pattern 
symmetric about the center for added throughput, coming to $\sim$25$\%$ 
compared to a completely open pupil.  The azimuthal coverage is $\sim$3 
times better at a $z$=10$^{-6}$ than the single opening design, and
suffers two orders of magnitude degradation in contrast.

\section{Fabricating a GAPM}
\label{fab}
Once a design was chosen the masks were fabricated.  For the initial observing
 run
at Mt. Wilson, we chose to have the masks made by Photo-Chemical Machining
(PCM).  This technique has been used to produce masks to block thermal radiation
 from telescope structures for near-IR observing and for creating Lyot stops.
 
The process of PCM, also called Photo-Etching or Photo-Chemical Milling, 
involves using a thin metal sheet that is coated with a light sensitive polymer.  
Then a UV photo imaging tool is used to imprint the desired design on the
sheet.  It is then developed much like film and chemically etched by an
aqueous solution of ferric chloride (FeCl$_3$).

Several masks are present in the PIRIS camera mainly for the traditional lyot
coronagraphic modes.  They were fabricated by Newcut, Inc. (Newark, NJ).  
For the GAPMs we submitted CAD designs based on the simulations performed
to Newcut and they fabricated the masks.  A sheet of 25-50 masks 
with a diameter of $\sim$ 4 mm were 
fabricated at very low cost within a few weeks.  When they were delivered
they were photographed on an optical telescope with 5x and 50x magnification.

This technique can provide the basic shape we need, but 
has difficulty preserving the exact shape of the design in the smallest 
regions.  The edges of the gaussians on the mask were truncated well before 
they 
mathmatically would be.  Variations on the order of 10$\mu m$ are also observed
in the masks.  Both of these imperfections can degrade contrast, which is 
discussed further in Section \ref{test}.  These imperfections are likely caused
by the photo printing as well as the chemical etching.  for instance, the
corner truncation and width variation can be caused by underexposure of the 
light sensitive polymer.  The rugged edge can be caused by non-uniform chemical
etching.  Since the etching is isotopic, local changes in the physical and 
chemical conditions of the etchant cause irregularities.

\section{Testing the Prototype}
\label{test}
We placed the fabricated pupil mask on the Penn State Near-IR Imager and 
Spectrograph (PIRIS) and ran tests both in the lab and on the 100'' Mt. Wilson
telescope.  We used the prototype as part of a survey for faint companions
 around nearby solar type 
stars \cite{chakraborty02,debes02}.
The stars $\epsilon$ Eridani (GJ 144 = HD 22049 = HR 1084, V=3.72, d=3.2 pc)
and $\mu$ Her A (GJ 695A = HD 161797 = HR 6623, V=3.41, d=8.4 pc) were observed
 with the GAPM.
The star $\epsilon$ Eridani was observed with seven 4s integrations in the K band,
while $\mu$ Her A was observed with one 3s and one 10s integration in the H
band.  Both objects were also observed in normal imaging modes.  The plate
scale and orientation for PIRIS was determined by measuring
the positions of several stars in the Orion Nebula and comparing them to 2MASS
 data.  We find that the plate
scale of PIRIS is $.082^{\prime\prime} \pm .001^{\prime\prime}$
pixel$^{-1}$.

The GAPM was aligned 23$^{\circ}$ clockwise from North to line up
with the support structure of the telescope.  Therefore, the axis of greatest
contrast would be oriented perpendicular to the alignment of the mask or along
a line running from the NW to the SE in an image.

For our coronagraphic modes we used a stop in the focal plane that has a
gaussian transmission function \cite{nakajima94} with a fwhm of
$\sim$1.1$^{\prime\prime}$ in an image.
A Lyot stop was placed in the pupil
 plane whose dimensions were chosen for an optimal combination of throughput
and contrast (See Ref ~\citenum{debes02}).

\begin{figure}
   \begin{center}
   \begin{tabular}{c c}
   \includegraphics[height=5cm]{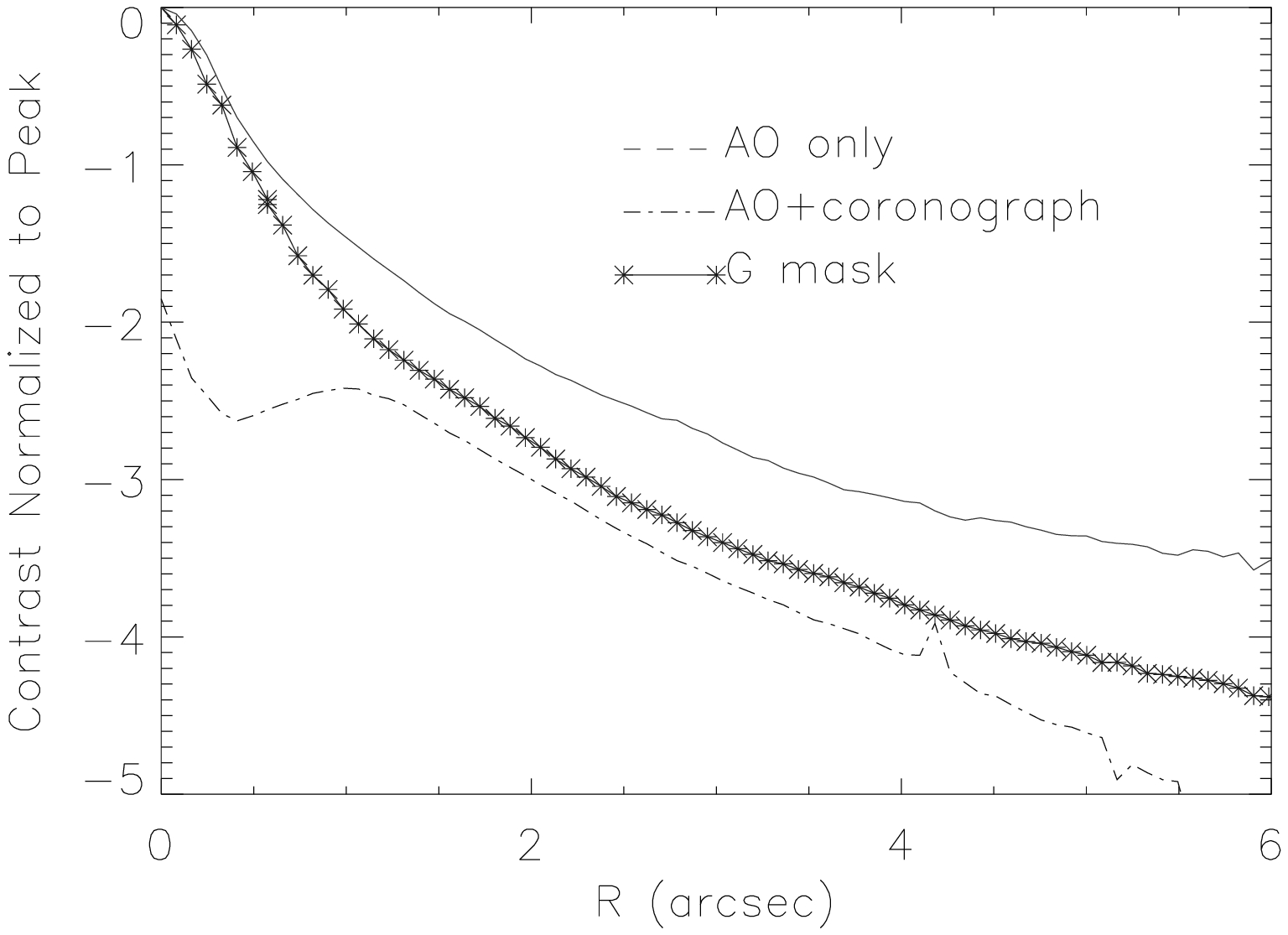} &
   \includegraphics[height=5cm]{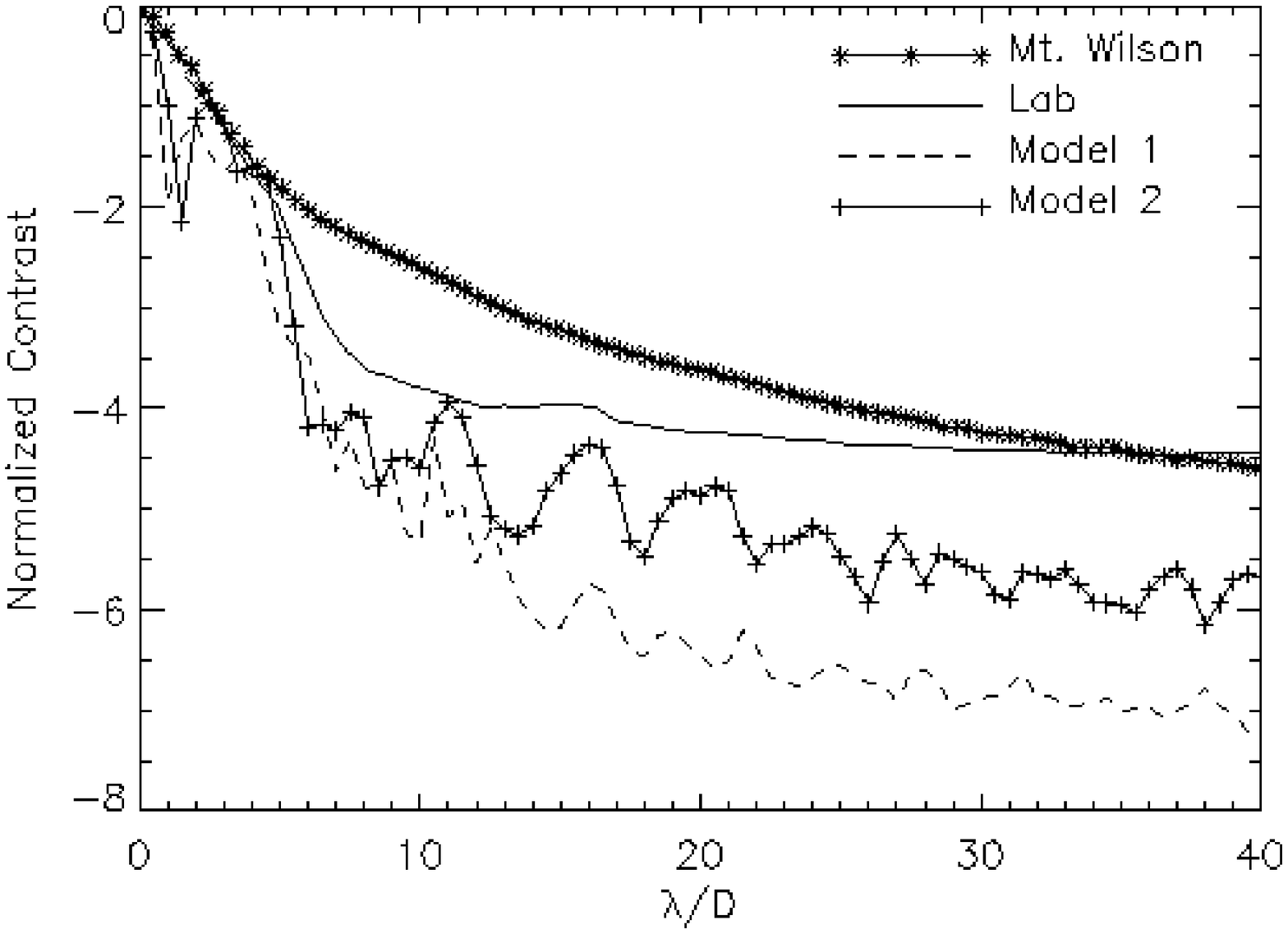}
   \end{tabular}
   \end{center}
   \caption[example]
{ \label{fig7} (left) Azimuthally averaged PSF profiles for three different
modes observed with on the Mt. Wilson telescope. (right) Azimuthally averaged
PSF profiles for the GAPM multiaperture design tested at Mt. Wilson, along with
results of a lab test and two different models.
}
   \end{figure}

In order to test the level of contrast achieved by the gaussian mask in
comparison to other modes of PIRIS, coronographic and adaptive optics
observations were taken of the star+faint companion system Gl 105 \cite{gol95,gol00}.  Figure
\ref{fig7} shows the differences in the three modes.  The observations 
were azimuthally averaged in
high contrast areas.
For Gl 105, we azimuthally averaged the PSF with the exception of 35$^\circ$
on either side of the faint companion.  The resulting
profiles were then normalized to peak flux.  Further scaling had to be
performed for the coronagraphic and gaussian pupil mask profiles.  In these two
cases the peak flux could not be determined due to the coronagraphic mask for
Gl 105 and due to saturation for $\epsilon$ Eridani.  In the case of Gl 105
this problem was circumvented by the presence of the companion which was used
to properly scale the flux.  The error in the curve for the coronograph is due
mostly to uncertainties in the positions  and flux ratio of the companion and the central
star, but
should be no larger than on the order of 10\%.
  In the case of $\epsilon$ Eridani we took
unsaturated images of a dimmer standard star and normalized the azimuthally
 averaged PSF at .82$^{\prime\prime}$.  The profile of the GAPM is
a composite of those two sets of observations.
The error in this normalization is dominated by variations in the PSF and to
 estimate the
effects of
this we varied the point at which we normalized the two profiles.  We found 
that at most the error
introduced is of the order 25\%.

As can be seen in Figure \ref{fig7}, the GAPM
performs better than AO alone and is $\sim$2 times worse than a traditional
coronograph $>$ 1$^{\prime\prime}$.  This is without any attempt to
block the light of the central star.  

Lab testing was also performed in the J band.  The setup involved taking an
incandescent lamp and simulating a point source to sample the PSF generated by
the different masks.  An optical fiber took light
from an incandescent lamp where it passed through a micro objective and a 
pinhole.  The light then was collimated by a collimator
achromat.  After the collimator it was focused onto the slit wheel 
aperture by an image achromat.
The image achromat 
also forms
an exit pupil, ~ 1.9 m away from the focal plane, mimicking the Mt. Wilson 
100inch
exit pupil.  On the slit wheel we placed our focal plane coronagraphic masks.
The light then travels through the camera optics of PIRIS where it is read
by the 256 $\times$ 256 PICNIC array.

Figure \ref{fig7} shows an azimuthally averaged comparison 
between the data taken at Mt. Wilson,
lab tests, and two theoretical simulations as a function of $\lambda/D$.  
One simulation, called model 1, represents a completely ideal situation where
the mask is perfectly created and no wavefront errors exist.
The second simulation, model 2, takes
the observed shape of the masks under magnification as the apertures and 
neglects other errors.  All of the data and simulations
were averaged over the same angle.  One can see that the theoretical
simulation of the observed shape matches the lab data quite well, off by less 
than an order of magnitude close to the center.  The observed shape errors also
degrade the contrast achievable by the idealized design.  Finally the effect
of the atmosphere is present in the Mt. Wilson data, which further
degrades the contrast from what is possible.  
  
From the azimuthally averaged profiles, limits to the type of companions we
could have detected around $\epsilon$ Eridani can be estimated.  Beyond 4$^{\prime\prime}$
 ($\sim$ 13 AU), any companion with $\Delta m_K$ $<$ 9.3 mag would
have been detected, and at 8$^{\prime\prime}$ ($\sim$ 26 AU) any companion
$<$ 11.8 mag would have been detected.  These translate to $M_K$ of 13.6 and
16.1 respectively. We looked at
the models of Ref.~\citenum{burrows97} and Ref.~\citenum{chabrier00} to determine the range
of possible companion masses that we could have detected based on .7 Gyr and
M$_K$=13.6 and 16.1, corresponding to 38$\pm$8 M$_J$ and 20$\pm$5 M$_J$.  We
derived these estimates by taking a value intermediate to the two models,
while the error represents their spread.  These results are reported in more
detail in Ref.~\citenum{debes02}.

\begin{figure}
   \begin{center}
   \begin{tabular}{c}
   \includegraphics[height=7cm]{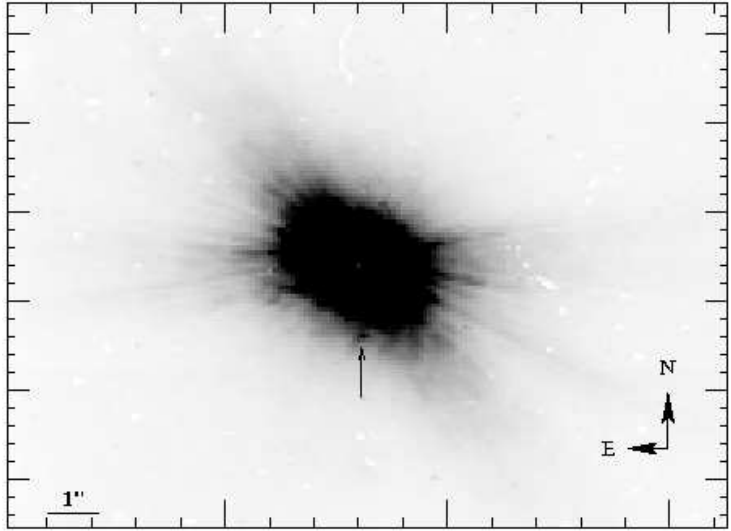} 
   \end{tabular}
   \end{center}
   \caption[example]
{ \label{fig8} Raw Image of Mu Her A.
}
   \end{figure}

A faint companion around $\mu$ Her A was detected with high S/N in both
GAPM and AO images.  A raw image of 
This companion was previously $\mu$ Her A and its companion is shown in Figure
\ref{fig8}
detected in R and I band AO images at Mt. Wilson, and used to explain a radial
 velocity acceleration corresponding to a $\sim$30 yr
orbit \cite{turner01,cumming99,cochran94}.  

  Our observations confirm
 the dimmer object to be a proper motion
companion and thus physically bound to the brighter star \cite{debes02}.  
H and K 
adaptive 
optics images of $\mu$ Her A were also performed.  Unfortunately,
standard star measurements with the GAPM could not be performed, but we
took H and K photometry with the normal imaging mode of PIRIS.  We find that
 for the companion
 $m_H$=8.5 $\pm$ .1 and $m_K$=7.9 $\pm$ .1, which corresponds to $M_H$=8.9 and
 M$_K$=8.3.  We used the models of Ref.~\citenum{baraffe98,baraffe02} to conclude that
its mass should be $\sim$.13 M$_\odot$.  For a more detailed description of 
our results refer to Ref.~\citenum{debes02}.

\section{Equal Throughput Comparisons}
\label{equal}
\begin{figure}
   \begin{center}
   \includegraphics[height=8cm]{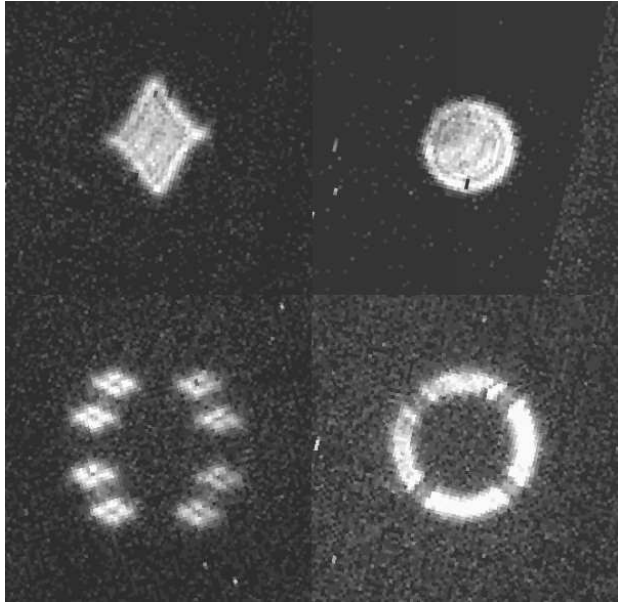} 
   \end{center}
   \caption[example]
{ \label{fig9} Images of the different pupils used for the equal throughput
comparisons.
}
   \end{figure}

A fairer comparison between a Lyot coronagraph and the GAPM is to have equal 
throughput designs and compare their contrast levels.  As part of  
lab experiments that we are performing we compared two new GAPM designs with 
a Lyot coronograph that had a comparable throughput.  Two types of designs 
were tested, idealized apertures with no secondary structure and realistic
masks that will be used for future observing.  Figure \ref{fig9} shows the 
four types of masks used, a single gaussian opening accompanied by an ideal 
Lyot coronagraph undersized by the prescription given in Ref.~\citenum{sivar01}
to give 20$\%$ throughput, the same as the GAPM.  The multi-aperture 
GAPM has $\sim$30$\%$ throughput as well as the corresponding Lyot stop.
Unsaturated images GAPMs were taken to predict the flux
for the longer, saturated images.  This predicted flux was used
to estimate the peak flux for the longer exposures taken.  The critical fact
of these experiments is to get enough counts to make sure that the PSF 
dominates the read noise.  We took exposures that had on the order of 10$^{7}$
counts.  We found that this was insufficient to get high S/N on the fainter 
portions of the PSF and we estimate that beyond 1-2$^{\prime\prime}$ the read
noise begins to dominate.  Longer integrations are planned.  

For the 
coronagraphic modes we used a gaussian transmission focal plane mask with a
FWHM of 500$\mu m$.  In
the J band this corresponded to a mask with a FWHM of $\sim 11\lambda/D$.
Short exposures were taken without the mask for estimating the peak flux of an
unblocked point source for a given exposure time.  The mask was then carefully
aligned to within 1 pixel to block the point source.  

\begin{figure}
   \begin{center}
   \begin{tabular}{c c}
   \includegraphics[height=6cm]{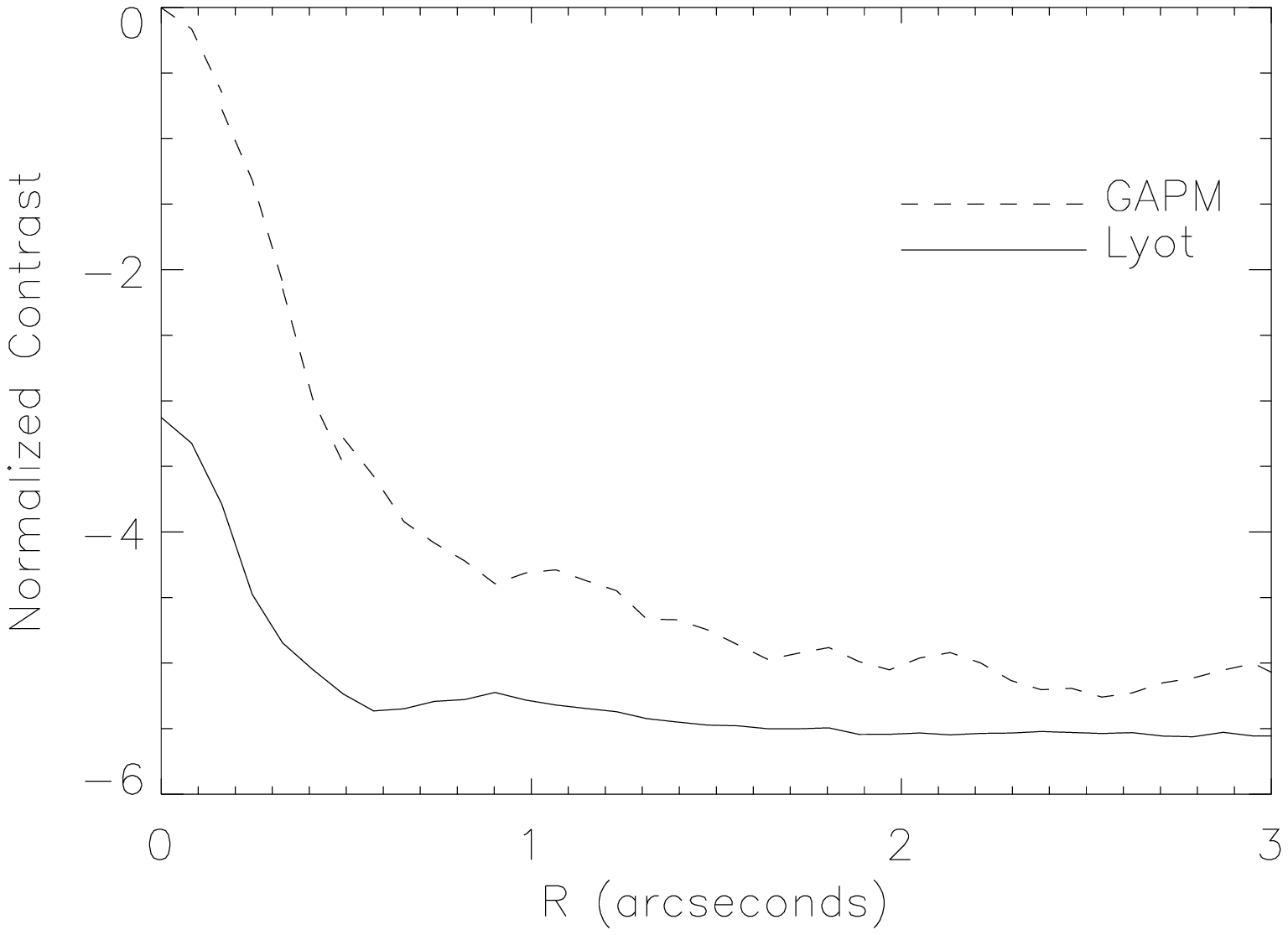} &
   \includegraphics[height=6cm]{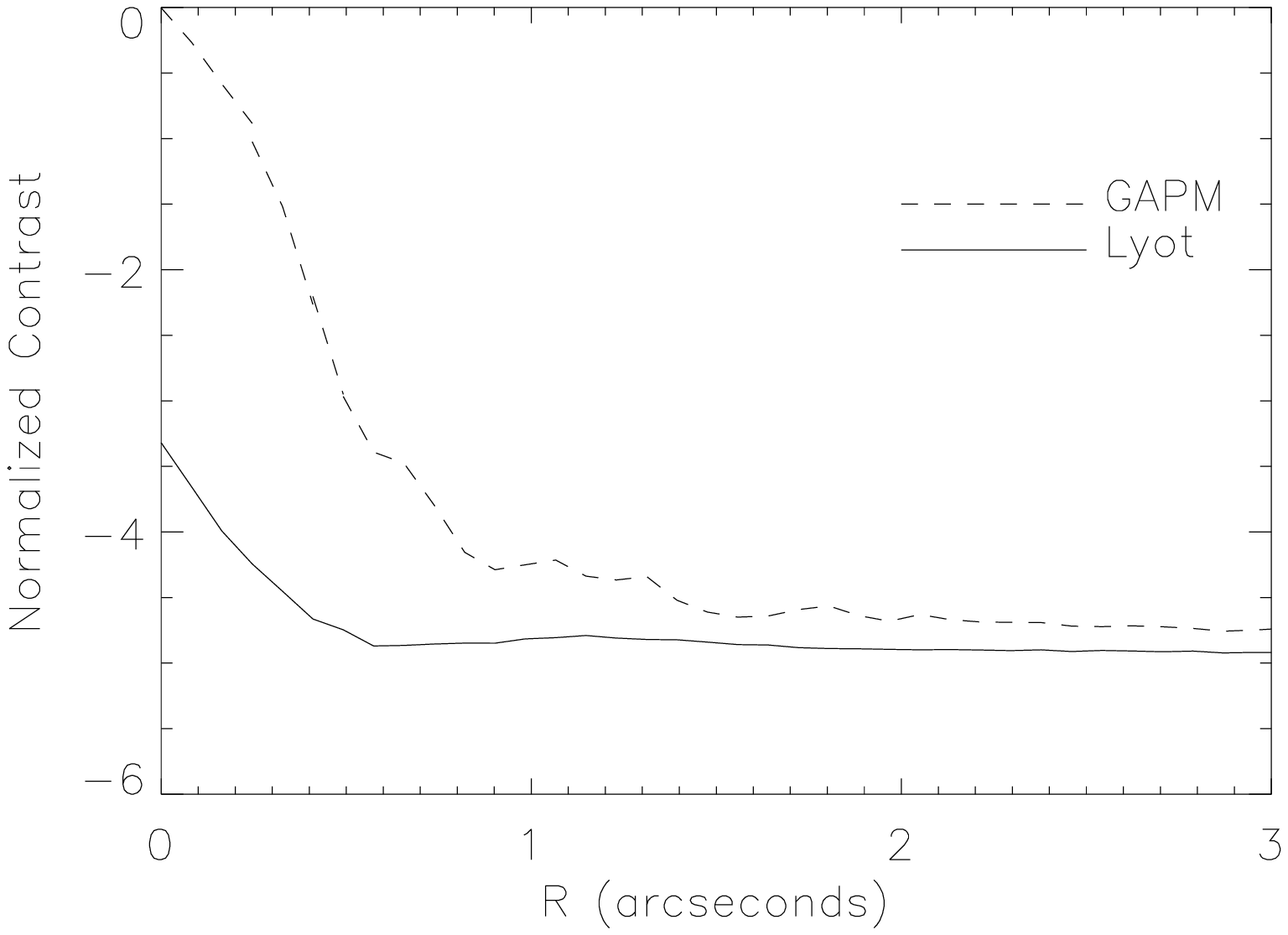}
   \end{tabular}
   \end{center}
   \caption[example]
{ \label{fig10} Equal throughput comparisons of GAPMs and a Lyot coronagraphs.
(left) Azimuthal average of single aperture and idealized Lyot.  (right) 
Azimuthal average of multi-aperture and realistic Lyot. 
}
   \end{figure}

Figure \ref{fig10} shows the results of these lab tests, which represents the
best performance of our GAPMs and coronagraphic modes.  At $>$ 10$\lambda/D$
the GAPMs perform at $z=3 \times 10^{-5}$ for the idealized version and 
  $z=6 \times 10^{-5}$ for the multi-aperture design.  The coronagraphs 
perform better, both achieving less than 10$^{-5}$ contrast.  
 
\section{Conclusions}
\label{concl}
We have performed several simulations, lab tests, and telescope observations 
with GAPMs and Lyot coronagraphs in order to better understand
the interplay between theory and the reality of observations.  GAPMs alone 
provide an improvement over a simple circular aperture for quick high contrast
imaging.  They are very sensitive to an accurate reproduction of shape and thus
need accuracies that may be as restrictive as sub-micron precision.  This is 
possible with new nanofabrication techniques that have been perfected at the
Penn State Nanofabrication facility, where future masks may be produced. 
Precisely fabricating these masks can potentially improve performance to 
the ideal limit for a mask provided it is above the scattered light
limit of the telescope, bringing it in line with Lyot coronagraphs
 of comparable throughput.  Demonstration of these masks in conjunction with
 an active optics system 
could present a workable example of a quick way to survey for faint companions
without needing to
incur the overhead cost of presice alignment behind a mask.
  GAPM performance on the ground
will ultimately be limited by the ability of adaptive optics systems to 
supress atmospheric turbulence.
  
\acknowledgements

The authors would like to acknowledge D. McCarthy for loaning part of the optics for PIRIS, R. Brown (Colorado) for
the PICNIC array, C. Ftaclas for coronagraphic masks, and A. Kutyrev for
filters.  Several important discussions with D. Spergel, M. Kuchner,
W. Traub, C. Burrows, and R. Brown (STScI) were crucial in
our understanding of gaussian apertures, band limited masks, and HST image
performance.
We would also
like to thank the invaluable help of the
the Mt. Wilson staff and L. Engel for design help with the GAPM.

J.D acknowledges funding by a NASA GSRP fellowship under grant NGT5-119. This
work was supported by NASA with grants NAG5-10617, NAG5-11427 as well as the
Penn State Eberly College of Science.

\bibliography{spie}
\bibliographystyle{spiebib}   

\end{document}